# Cascadable in-memory computing based on symmetric writing and read out

Lizheng Wang[1†], Junlin Xiong[1†], Bin Cheng[2]*, Yudi Dai[1], Fuyi Wang[1], Chen Pan[2], Tianjun Cao[1], Xiaowei Liu[1], Pengfei Wang[1], Moyu Chen[1], Shengnan Yan[1], Zenglin Liu[1], Jingjing Xiao[3], Xianghan Xu[4], Zhenlin Wang[1], Youguo Shi[3], Sang-Wook Cheong[4], Haijun Zhang[1], Shi-Jun Liang[1]*, Feng Miao[1]*

[1] National Laboratory of Solid State Microstructures, Institute of Brain-Inspired Intelligence, School of Physics, Collaborative Innovation Center of Advanced Microstructures, Nanjing University, Nanjing 210093, China.

[2] Institute of Interdisciplinary Physical Sciences, School of Science, Nanjing University of Science and Technology, Nanjing 210094, China.

[3] Institute of Physics, Chinese Academy of Sciences, Beijing 100190, China.

[4] Center for Quantum Materials Synthesis, and Department of Physics and Astronomy, Rutgers, The State University of New Jersey, Piscataway, NJ, 08854, USA

† Contribute equally to this work.

* Correspondence Email:

bincheng@njust.edu.cn; sjliang@nju.edu.cn; miao@nju.edu.cn

## Abstract

The building block of in-memory computing with spintronic devices is mainly based on the magnetic tunnel junction with perpendicular interfacial anisotropy (p-MTJ). The resulting asymmetric write and read-out operations impose challenges in downscaling and direct cascadability of p-MTJ devices. Here, we propose that a new symmetric write and read-out mechanism can be realized in perpendicular-anisotropy spin-orbit (PASO) quantum materials based on $Fe_3GeTe_2$ and $WTe_2$. We demonstrate that field-free and deterministic reversal of the perpendicular magnetization can be achieved by employing unconventional charge to z-spin conversion. The resulting magnetic state can be readily probed with its intrinsic inverse process, i.e., z-spin to charge conversion. Using the PASO quantum material as a fundamental building block, we implement the functionally complete set of logic-in-memory operations and a more complex nonvolatile half-adder logic function. Our work highlights the potential of PASO quantum materials for the development of scalable energy-efficient and ultrafast spintronic computing.

## INTRODUCTION

A magnetic tunnel junction with perpendicular interfacial anisotropy (p-MTJ), composed of a ferromagnetic metal/tunnelling oxide/ferromagnetic metal heterostructure (*1, 2*), is the leading candidate for nonvolatile spin memory and logic applications (*3-6*). The information is encoded as the perpendicular magnetization state of the magnet in the p-MTJ and is electrically written and read out with distinct mechanisms. The write operation of the magnetization is conventionally achieved by spin-transfer torque (*7*) or spin-orbit torque (*8-10*) with the assistance of an in-plane magnetic field, while the read out of the resulting magnetic state is enabled by measuring the tunnelling magnetoresistance. With the increasing demand for energy-efficient and scalable spin logic circuits, symmetric field-free write operation and reliable read-out operation of directly driven next-stage circuits must be achieved, which has been the major hurdle for conventional p-MTJ devices (*5, 11, 12*). Exploiting novel physical properties of artificially engineered quantum materials with the coexistence of perpendicular magnetism and unconventional spin–orbit coupling (SOC) effects may offer a promising approach to realize alternative spin devices beyond conventional p-MTJ structures to address these challenges.

Emergent quantum materials comprising a van der Waals (vdW) magnet and a topological semimetal could provide unique opportunities for advancing ultralow-power spintronic applications (*13-19*). Different from the interfacial anisotropy in the conventional p-MTJ, recently discovered layered magnets with intrinsic large perpendicular magnetocrystalline anisotropy can be readily tuned by various external knobs, such as strain (*2, 20, 21*) and electrical gating (*22-24*). In addition, the nontrivial distribution of the Berry curvature and spin texture present in topological semimetals with strong SOC shows great promise in realizing unconventional and highly efficient charge-spin conversion (*25-30*). Employing a Lego-like fashion enables artificial synthesis of perpendicular-anisotropy spin-orbit (PASO) quantum materials and achievement of symmetric write and read-out operations of the perpendicular magnetization by using unconventional charge to z-spin conversion and its intrinsic inverse effect, as schematically shown in Fig. 1. For the write operation (left panel in Fig. 1), flowing an electron current in a given in-plane direction can generate an out-of-plane spin polarization that would diffuse into the ferromagnetic layer and induce an unconventional out-of-plane spin-orbit torque (*17, 18, 31*). This unconventional spin-orbit torque is able to switch the perpendicular magnetization from the UP state to the DOWN state without the assistance of an external field. For the read-out operation (right panel in Fig. 1), the vertically electron current is polarized with the resulting magnetization, and the spin polarization injected across the interface can be converted into an in-plane charge current, which can be directly used for driving the next-stage spintronic circuit. Conversely, the write operation that switches the magnetization from the DOWN state to the UP state can be realized by reversing the direction of the charge current, with the corresponding read-out operation implemented by injecting the reversed spin polarization. In addition, employing the Lego-like fashion in the design of spintronic devices allows for creating atomically flat interfaces and thus exhibits huge potential in solving the challenges associated with interface quality in conventional MRAM devices, such as increasing tunneling magnetoresistance (*32*), raising the operating temperature (*33*) and so on.

## RESULTS

In this work, a PASO quantum material was artificially synthesized by stacking topological semimetal WTe$_2$ onto layered ferromagnetic material Fe$_3$GeTe$_2$ (FGT). FGT is a layered vdW ferromagnetic metal with perpendicular magnetic anisotropy (PMA) and holds great promise for high-density spin device applications (*21, 24, 34, 35*). With strong SOC and low crystalline symmetry (*27, 36, 37*), WTe$_2$ can generate z-spin polarization ($\sigma_z$) when flowing a charge current along the low-symmetry axis of WTe$_2$ (*25*), which can diffuse into FGT and may thus facilitate field-free switching of the perpendicular magnetization of FGT. In contrast, when a charge current is applied along the high-symmetry axis of WTe$_2$, the mirror symmetry in the *bc* plane requires that $\sigma_z = 0$. In addition, the conductivity of WTe$_2$ is much larger than that of FGT in the PASO material (see Fig. S1), leading to most of the charge current being shunted through WTe$_2$ and boosting energy-efficient magnetization switching. Fig. 2A and Fig. 2B show a schematic and corresponding optical images of the PASO material-based device, with the WTe$_2$ and FGT flakes approximately 29.9 nm and 10.2 nm in thickness, respectively (see the atomic force microscope images and corresponding height profiles in Fig. S2). The crystalline axis of WTe$_2$ was determined by polarized Raman spectroscopy (Fig. S3) to ensure that the charge current is injected along the low-symmetry axis of WTe$_2$ (i.e., the x-axis in Fig. 2A). The perpendicular magnetization of FGT was characterized by monitoring the anomalous Hall resistance of the device, with the results at different temperatures shown in Fig. 2C. The square-shaped hysteresis loop appearing at 50K indicates a single-domain ferromagnetic phase of the PASO material. With increasing temperature, the hysteresis loop gradually disappears and eventually vanishes at ~190 K, which is consistent with the reported Curie temperature of FGT flakes (*21, 34*).

**Current-induced switching of perpendicular magnetization**

We then examined the current-induced magnetization switching in the PASO device by sweeping the pulsed current $I_{\text{pulse}}$ at 120 K and monitoring the change in the anomalous Hall resistance at different external magnetic fields $\mu_0 H_x$. Here, $\mu_0 H_x$ and $I_{\text{pulse}}$ were both applied parallel to the low-symmetry axis of the WTe$_2$ flake, i.e., the x-axis indicated in Fig. 2A. By applying the in-plane magnetic field, we investigated the mechanism of field-free switching, especially demonstrating the role of y and z components of current-induced spin polarization play on the field-free switching in our device. When the external field is applied in the opposite direction to the x-axis (i.e., $\mu_0 H_x$ = -20 mT), sweeping $I_{\text{pulse}}$ upwards or downwards to a critical value (corresponding to a current density of $5.0 \times 10^6$ A cm$^{-2}$) reverses the magnetization of FGT and give rise to a clockwise switching polarity (Fig. 2D). Notably, increasing the current density to approximately $5.9 \times 10^6$ A cm$^{-2}$ allows realization of field-free magnetization switching with the same switching polarity as above. This clockwise switching polarity is retained at $\mu_0 H_x$=20 mT and eventually reversed at $\mu_0 H_x$ =40 mT. These results indicate that a nonzero critical in-plane field (labelled $\mu_0 H_C$, with 20 mT < $\mu_0 H_C$ < 40 mT) exists, below and above which the switching polarities of the magnetization are distinct in the PASO device. It is the fact that the out-of-plane torque induced by the z-spin is larger than that induced by in-plane magnetic field of 20 mT, and smaller than that of 40 mT. Notably, no deterministic switching can be realized when the current flowing along the high-symmtery axis, again indicating the absence of current-induced z-spin polarization in this case (Fig. S4).

We attribute the appearance of the nonzero $\mu_0 H_C$ in the PASO device to the existence of an additional out-of-plane damping-like torque $\mathbf{T_z^{DL}} = \mathrm{H_z^{DL}} \mathbf{m} \times (\mathbf{m} \times \boldsymbol{\sigma})$, which arises from the current-induced z-spin polarization (i.e., $\boldsymbol{\sigma} = \mathbf{z}$) (see detailed analysis in Methods and the micromagnetic simulation and corresponding analysis of magnetization switching in Fig. S5). The existence of this unconventional charge to z-spin conversion in the PASO device can be further verified by the existence of a z-spin generated field-like effective field $\mathbf{H_z^{FL}} = H_z^{FL}\boldsymbol{\sigma}$ which is proportional to the applied charge current (see a detailed analysis in Methods). Here, the $H_z^{FL}$ manifested as a current-dependent perpendicular field ($H_{eff}^z$) was characterized by measuring the anomalous Hall resistance as a function of the perpendicular magnetic field under different direct currents ($I_{dc}$) (*9*) (Fig. S6). This unusual current induced z-spin may arise from longitudinal planar spin Hall effect (p-SHE), which converts the in-plane charge current into out-of-plane spin current with collinear spin polarization, and could exist in WTe$_2$ due to its nontrivial distribution of Berry curvature and spin texture in momentum space (*38-40*). The theoretically calculated spin Hall conductivity $\sigma_{zx}^z$ of WTe$_2$ is non-zero, further indicating the existence of p-SHE in our device (see Fig. S7 and Methods for calculations details). In addition, the two-fold screw rotation and glide mirror symmetries of WTe$_2$ are broken at the WTe$_2$/FGT interface; as a result, the Rashba-Edelstein effect may also contribute to the current-induced z-spin (*17, 31, 41*), which requires more theoretical investigations to elucidate in the future.

**Symmetric read out of the perpendicular magnetization state**

Note that such unconventional charge-to-spin conversion in the PASO based device allows a symmetric read-out operation by using its inverse process, i.e., converting the out-of-plane spin polarization into a transverse charge current. Fig. 3A shows a schematic and optical image of the PASO device for reading out the data encoded in the spin polarization state. When a charge current ($I_{supply}$) is perpendicularly injected into the PASO device, the electrons with spin polarization aligned with that in the FGT layer will be filtered to flow through the WTe$_2$ layer. The resulting spin polarization injected across the interface is then converted into a transverse charge current in WTe$_2$, which can be detected as a transverse voltage $V_{read}$ (inset of Fig. 3B). As the magnetic state is switched by the perpendicular magnetic field, a sharp jump occurs in the Hall resistance $R_{Read}$ (Fig. 3B), which is defined by $R_{read} = V_{read}/I_{supply}$. Here, $I_{supply}$ is the absolute value of applied current. When a reverse charge current is injected into the PASO device, the direction of the generated transverse charge current output will also be reversed (inset of Fig. 3C), thus flipping the switching polarity of the read-out signal of the magnetic state (Fig. 3C). In addition, the voltage change ($\Delta V$) (defined as the difference in $V_{read}$ between two different magnetization configurations of FGT, see Fig. S8) and $H_{eff}^z$ (given in Fig. S6) are both proportional to the applied charge current. This similarity strongly suggests that the spin-charge interconversion in the write and read-out operations of perpendicular magnetization arises from the same physical origin. It is noted that such jump of read out signal is absent when the voltage is probed along the high-symmetry axis (Fig. S9). Compared to the read-out operation of the in-plane magnetic state, e.g., using the conventional inverse SHE in CoFe/Pt devices (*12*), our strategy of reading out perpendicular magnetization by adopting the intrinsic z-spin to charge conversion not only enables to directly drive the next-stage device, but also raises the conversion efficiency at the same scaling size by one order of magnitude (see calculation details in Methods).

## Cascaded field-free spin-orbit logic in memory

The symmetric field-free write and directly driven next-stage-circuit read-out operations allow the use of PASO quantum materials as a building block for implementing ultrafast, energy-efficient and cascadable field-free spin-orbit logic-in-memory circuits. Fig. 4A demonstrates a schematic of the PASO material-based spin-orbit logic-in-memory device with three separate input terminals (see basic characterizations of the devices in Fig. S10). Two input signals (represented by $I_A$ and $I_B$) and one control signal (represented by $I_{control}$) are fed into the WTe$_2$ layer, which are used for deterministic reversal of the perpendicular magnetization in the FGT flake via the current-induced z-spin polarization in the WTe$_2$ flake. As such, field-free deterministic switching can be used to implement different logic functions (*i.e.*, NAND, NOT, NOR) in the individual device (Fig. 4B). The outputs of the logic gates are determined by the memory states of the device, except for the NOT gate, where $I_A$ and $I_B$ are combined and fed into the device as an input signal. To illustrate how the memory states affect the logic outputs, we present the operating mechanisms for distinct logic gates in Fig. S11. For the NAND gate, if the initial perpendicular magnetization points upward (denoted as memory state "1"), two input currents both flowing from the left toward right (i.e. "11" as input signals) in the WTe$_2$ film would generate spin current that diffuses and accumulates at the interface. The spin polarization of pointing downward would exert a torque on the perpendicular magnetization in the FGT film and switch it from upward to the downward (corresponding to memory state "0"). Changing direction of two currents flowing ("00") would cause the accumulation of spin polarization pointing upward at the interface and the resulting torque has a direction in parallel with the perpendicular magnetization. As a result, the initial perpendicular magnetization is retained. When the two currents are flowed in antiparallel (i.e. "01" or "10" as input signals), the effective out-of-plane torque exerted on the perpendicular magnetization of the FGT is zero, thus the direction of the perpendicular magnetization is kept unchanged. To implement the NOR gate, $I_{control}$ is first used to program the initial memory state "1" to the memory state "0". All the input current signals would not cause the reversal of the perpendicular magnetization except for both of currents flowing from right to left, where the perpendicular magnetization is changed from downward into upward by the spin torque. Different from the NAND and NOR gates, the NOT gate can be implemented by flowing two currents along the same direction. Note that the implementation of NOT gate is independent of the initial memory state. To demonstrate the proof of concept, we set the positive current pulse (+8 mA) as input logic state '1' and the negative current pulse (-8 mA) as input logic state '0' in the experiments. The logic output is represented by the final state of the perpendicular magnetization, which is characterized by the anomalous Hall resistance. Fig. 4C presents an experimental demonstration of the NAND logic gate, in which a control signal of -16 mA is employed to set the initial magnetization state to the UP state (denoted as memory state "1") in a deterministic manner before feeding the input current signal. By reversing the direction of the control current signal, we can implement the NOR gate, with the experimental results shown in Fig. 4D. For experimental implementation of the NOT gate, current pulses of the same polarity are alternatively applied to the device without a control current signal (Fig. 4E). By employing the read-out operation able to generate a charge current output, which directly drives the field-free magnetization switching of another PASO device, these three logic gates can be cascaded to realize arbitrary complex in-memory logic functions. Fig. 4F illustrates the schematic circuit of the half-adder based on four PASO devices, which consists of XOR and AND gates to generate SUM and CARRY signals, respectively. In the circuit, $I_A$ and $I_B$ are used as the input signals, and the interconversion signals ($I_{supply}$) are employed to read out the preceding logic stage as the charge current (whose sign depends on the magnetic state) to drive the next logic stage. Fig. 4G shows the logic operations of the half-adder and the operating waveforms used for control and interconversion signals of the different devices. When feeding two signals of "1" and "1" into Device 1 and Device 2, the output logic states "0" and "1" are generated from Device 3 (as an AND gate) and Device 4 (as an XOR gate), respectively (see more details of the logic operations in the half-adder in Methods and Fig. S12).

## DISCUSSION

In conclusion, by using the unconventional charge to z-spin conversion and its symmetric inverse effect, PASO quantum materials can realize the symmetric field-free write and directly driven next-stage-circuit read-out operation. This symmetric write and read-out operation enables to overcome the challenges associated with asymmetric write and read-out operations as well as the failure to directly generate a current or a voltage in the conventional p-MTJ-based device. Notably, the PASO-based device can be operated at ultralow power, and the energy consumption per switching event can potentially be reduced to 10 aJ by artificially synthesizing PASO quantum materials with larger charge-to-spin conversion efficiency ($> 1$) (*13, 42-44*) and conductivity ($> 10^6 /(\Omega \cdot m)$) (*30, 45*) (see more details in Supplementary Text). Moreover, a PASO material with a large PMA energy and an above-room-temperature Curie temperature is achievable (*33, 46, 47*) and shows potential in downscaling spin-orbit logic devices beyond sub-10 nm. These advantages that the proposed Lego-like PASO quantum materials may open an alternative avenue for ultrafast, highly energy-efficient and large-scale logic computing technology.

*Note added.*—In the peer review stage of our work, a related work by Kao et al., focusing on the field-free switching of the magnetization in FGT by unconventional spin–orbit torques in WTe$_2$, was published (*48*). In comparison, our work has proposed and experimentally demonstrated symmetric operations of write and read-out in a PASO device, and implemented cascadable and field-free logic-in-memory device and circuit.

## MATERIALS AND METHODS

### Device fabrication

The WTe$_2$ and FGT flakes were mechanically exfoliated onto the SiO$_2$/Si substrate in a glovebox filled with an inert atmosphere. The bottom electrodes (2 nm Ti/30 nm Au) and top electrodes (5 nm Ti/50 nm Au) were patterned using the standard electron beam lithography method and deposited by standard electron beam evaporation. Poly(propylene) carbonate (PPC) coated on polydimethylsiloxane (PDMS) was used to pick up the WTe$_2$ and FGT flakes and fabricate the PASO devices. The thickness of the WTe$_2$ and FGT flakes was characterized via a Bruker Multimode atomic force microscope.

### Characterizations

All the electrical measurements were performed in the Oxford cryostat with magnetic fields of up to 8 T and temperatures between 1.5 and 300 K. The magnetic field can be applied along the in-plane or out-of-plane direction of the device by rotating the sample stage. Electrical measurements for WTe$_2$/FGT heterostructures were performed by using lock-in amplifiers (Stanford SR830) and a Keithley 2636B dual-channel digital source meter. To characterize the magnetization state, the Hall resistance was measured by using a lock-in amplifier (SR830). In the measurement of switching the magnetization, large current pulses (write current, 200 μs) were first applied by using Keithley 2636B dual-channel digital source meter. After an interval of 5s, the Hall resistance was measured using a small alternating current excitation current of 5 μA. Local spin detection measurements for magnetic state read-out were carried out using a phase-sensitive lock-in amplifier.

## Calculation of the intrinsic SHC

We performed ab initio calculations by using the QUANTUM ESPRESSO package (QE) (*49*) to obtain the band structure of $WTe_2$ and the corresponding input of the WANNIER90 package (*50*). Next, we relaxed ions with the cell fixed to get a more adaptable structure before further calculations. Here, the plane-wave cutoff energy was 70.0 Ry ($\approx$ 952 eV) and a k-point grid of $9 \times 7 \times 5$ was used in the self-consistent calculation (*51*). The WANNIER90 package was used to get maximally localized Wannier functions and calculate the SHC. A dense k mesh of $430 \times 240 \times 100$ was employed to perform the Brillouin zone (BZ) integration for the intrinsic SHC, and a fixed refinement ($\approx$ 0.02 eV) was used to deal with the rapid variation of the spin Berry curvature in Brillouin zone. Fig. S7A-B shows the calculated electronic band structures projected by the spin Berry curvature (SBC) of the conventional SHC ($\sigma_{zx}^y$) and unconventional SHC ($\sigma_{zx}^z$), respectively. The theoretic results show that substantial SBC hotspots are concentrated on the bands close to the Fermi energy, which mainly contribute to the large SHC. We obtain $\sigma_{zx}^z$ by integrating SBC in the Brillouin zone for different Fermi energy, with results shown in Fig. S7C. In the vicinity of Fermi energy, the calculated positive $\sigma_{zx}^z$ indicates that the charge current flowing along x direction of $WTe_2$ flake would generate spin polarization with the out-of-plane component in the PASO material. This theoretical result is consistent with the experimental results presented in this work. With similar calculating method, we also calculated $\sigma_{zx}^y$ and found that the negative $\sigma_{zx}^y$ in the vicinity of the Fermi level is comparable with that reported in recent theoretical works (*28*) (Fig. S7C).

## Micromagnetic simulation

The dynamics of magnetization switching is described by Landau-lifshitz-Gilbert equation with an additional in-plane and out-of-plane damping-like torque term:

$$\frac{1+\alpha^2}{\gamma} \cdot \frac{d\mathbf{M}}{dt} = -\mathbf{M} \times \mathbf{H}_{\text{eff}} - \alpha \cdot \mathbf{M} \times (\mathbf{M} \times \mathbf{H}_{\text{eff}}) + \zeta_y^{DL} I_C \cdot \mathbf{M} \times (\hat{\mathbf{y}} \times \mathbf{M}) + \zeta_z^{DL} I_C \cdot \mathbf{M} \times (\hat{\mathbf{z}} \times \mathbf{M}),$$

where $\gamma$ is the gyromagnetic ratio, $\alpha$ is the Gilbert damping constant, $\mathbf{M}$ is the unit vector of magnetic moment, $\mathbf{H}_{\text{eff}} = H_{\text{ani}} M_z \hat{\mathbf{z}} + \mathbf{H}_{\text{ext}}$ is the sum of anisotropy field of FGT and external field $\mathbf{H}_{\text{ext}} = H_{\text{ext}} \hat{\mathbf{x}}$, $\zeta_y^{DL} I_C$ and $\zeta_z^{DL} I_C$ is the amplitude of the in-plane and out-of-plane damping-like filed, respectively, and $I_C$ is the magnitude of applied current pulse. In the simulation, we set $H_{\text{ani}} = 0.7$ T, $\alpha = 0.005$, $\zeta_y^{DL} = 12$ mT/mA, and $\zeta_z^{DL} = -2$ mT/mA under different external magnetic fields $H_{\text{ext}} = -0.1$, 0, 0.02 and 0.1 T. For $H_{\text{ext}} = -0.1$, 0 and 0.02 T, the initial state of $\mathbf{M}$ is set to $\hat{\mathbf{z}}$; while for $H_{\text{ext}} = 0.1$ T, the initial state of $\mathbf{M}$ is set to $-\hat{\mathbf{z}}$. We applied a positive current pulse (16 mA) with duration of 0.2 ns at the initial time, and then monitored the magnetization evolution based on the LLG equation (Fig. S5A).

## Dynamical model of the magnetization switching

In the simulation, we consider the main source of the torques exerting on the magnetization, including the precession torque ($\mathbf{T}_{\text{field}} \sim H_x \mathbf{m} \times \mathbf{x}$) induced by the external magnetic field, the in-plane damping-like torque ($\mathbf{T}_{DL}^y \sim \mathbf{m} \times (\mathbf{m} \times \mathbf{y})$) from in-plane spin polarization and out-of-plane damping-like torque ($\mathbf{T}_{DL}^z \sim \mathbf{m} \times (\mathbf{m} \times \mathbf{z})$) from out-of-plane spin polarization. The simulation results show that the dynamics of the magnetization switching can be qualitatively understood via two consecutive physics processes (see Fig. S5A). First, the conventional spin Hall effect would generate an in-plane damping-like torque on the perpendicular magnetization, which drives the magnetization to the plane with high energy efficiency and fast switching speed. Subsequently, the tilted magnetization is deterministically switched to the magnetization state corresponding to $\mathbf{M}_{\text{final}}$ by the out-of-plane torque.

To visualize the dynamics of magnetization switching in a step-by-step manner, we schematically show the corresponding physics pictures for different magnetic field in Fig. S5B, in which $\mathbf{M}$, $\mathbf{T_{field}}$ and $\mathbf{T_{DL}^z}$ represent the tilted magnetization, the out-of-plane torque generated by $\mu_0 H_x$ and the additional out-of-plane torque, respectively. When applying negative $\mu_0 H_x$, both $\mathbf{T_{DL}^z}$ and $\mathbf{T_{field}}$ point downward and push $\mathbf{M}$ to the same direction, leading to the clockwise switching polarity. In the case of $\mu_0 H_x = 0$, only $\mathbf{T_{DL}^z}$ exists and the switching polarity is unchanged. At the intermediate regime of $0 < \mu_0 H_x < \mu_0 H_C$, the magnitude of $\mathbf{T_{DL}^z}$ is larger than $\mathbf{T_{field}}$, and the switching polarity remains clockwise despite in an opposite orientation. In the regime of $\mu_0 H_x > \mu_0 H_C$, the tilted magnetization is eventually rotated to the direction of dominant $\mathbf{T_{field}}$, reversing the switching polarity to anti-clockwise.

**Analysis on the existence of out-of-plane torque**

In conventional NM/FM heterostructure-based SOT device, $\mu_0 H_x$ is necessary for generating an out-of-plane torque and thus enabling the deterministic switching of the perpendicular magnetization (*8*). Thereby, the reversal of magnetization switching polarity induced by current should occur at $\mu_0 H_x=0$, i.e., $\mu_0 H_C$ equals to zero. In our PASO device, the non-zero $\mu_0 H_C$ indicates an existence of additional out-of-plane torque ($\mathbf{T_{oop}}$) distinct from that generated by external $\mu_0 H_x$. To visualize the role of $\mathbf{T_{oop}}$ on non-zero $\mu_0 H_C$, we have schematically shown the corresponding dynamical model of the magnetization switching for different magnetic field above (Fig. S5B). From the analysis, we can infer that this torque $\mathbf{T_{oop}}$ with the form $\sim \mathbf{m} \times (\mathbf{m} \times \mathbf{z})$ is compensated by $\mathbf{T_{field}}$ from the external magnetic field equal to $\mu_0 H_C$, above and below which the switching polarity has opposite chirality. Moreover, the role of $\mathbf{T_{oop}}$ on the polarity of magnetization switching were further verified by the micromagnetic simulation of the magnetization switching (Fig. S5C). As such, we can attribute $\mathbf{T_{oop}}$ to the damping-like torque with the same form ($\mathbf{T_z^{DL}} = H_z^{DL}\mathbf{m} \times (\mathbf{m} \times \boldsymbol{\sigma})$) generated by current-induced z spin polarization ($\sigma = z$), consistent with the results reported in the previous works (*17, 18, 25*).

**The out-of-plane effective field generated by the z-spin polarization**

The existence of the current induced z-spin polarization can be further verified by investigating the current-dependent out-of-plane effective field. Fig. S6B shows the extracted values of $H_{eff}^z$ for different $I_{dc}$. The absence of $H_{eff}^z$ at $I_{dc} = 0$ indicates that the out-of-plane effective field arises from a pure current-induced effect, which can be attributed to the unconventional charge to z-spin conversion in the PASO device discussed below.

In the conventional heterostructure-based SOT device, the charge-current induced spin polarization would generate two types of effective fields, corresponding to the field-like torque ($\mathbf{T_z^{FL}} = \gamma H_z^{FL} \mathbf{m} \times \boldsymbol{\sigma}$) and the damping-like torque ($\mathbf{T_z^{DL}} = \gamma H_z^{DL} \mathbf{m} \times (\mathbf{m} \times \boldsymbol{\sigma})$), respectively (*31*). Here, $\mathbf{m}$ is the unit vector of magnetic moment which is parallel to the z-axis, and $\boldsymbol{\sigma}$ represents the orientation of current-induced spin polarization, which is reversed when applying the opposite charge current parallel to x-axis. Considering the mirror symmetry (*9*) with respect to the bc plane (Fig. S13A), the charge current flowing along the low-symmetry axis of $WTe_2$ flake could generate spin polarization with out-of-plane component, which is supported by the symmetry analysis and our theoretical calculations. The current-induced spin polarization would generate an effective field ($\mathbf{H_z^{FL}} = H_z^{FL}\boldsymbol{\sigma_z}$) consistent with the out-of-plane effective field shown in Fig. S6B. It is this effective field that displaces the perpendicular magnetization in the FGT flake out of its equilibrium alignment and induces the procession. The direction of $\mathbf{H_z^{FL}}$ is reversed with the opposite direction of charge current flowing under the operation of mirror symmetry, which is also consistent with the experimental results of current dependent out-of-plane effective field (Fig. S6B).

Besides, we can rule out the possibility that the out-of-plane effective field is generated by current-induced spin polarization along x- or y-axis. As illustrated in Fig. S13B, the conventional spin Hall effect or Rashba-Edelstein effect will convert the charge current into out-of-plane spin current with y-spin polarization (*52*). This y-spin polarization induced effective fields contain two terms, i.e. a field-like (FL) term $\mathbf{H_y^{FL}} = H_y^{FL}\boldsymbol{\sigma_y}$ pointing to y direction and a damping-like (DL) term $\mathbf{H_y^{DL}} = H_y^{DL}\mathbf{m} \times \boldsymbol{\sigma_y}$ pointing to x direction. Thus, these in-plane effective fields from the conventional SHE cannot explain the presence of $H_{eff}^z$. Similarly, we can exclude the possibility that the out-of-plane effective field results from other in-plane spin polarization generated by the conventional SHE.

The charge current induced spin polarization with the out-of-plane component allows for interpreting not only the current-dependence of the out-of-plane effective field observed in Fig. S6B, but also the existence of non-zero $\mu_0 H_C$ appearing in Fig. 2D. First, applying the charge current pulse leads to the switch of perpendicular magnetization towards y or -y direction by the y-spin polarization induced in-plane damping-like toque, depending on the direction of current flowing along the low-symmetry axis of WTe$_2$ flake. Next, the out-of-plane component of the resulting spin polarization would exert an out-of-plane damping-like torque $\mathbf{T_z^{DL}}$ on the in-plane magnetization (along -y or y direction). This torque is compensated by $\mathbf{T_{field}}$ from the critical magnetic field equal to $\mu_0 H_C$.

## Calculation of the conversion efficiency in the PASO device

When the charge current ($I_{supply}$) is vertically injected into the PASO device, the electrons with spin polarization will be aligned with the magnetization of FGT and flow through the WTe$_2$. Then, the resulting spin current is then converted into a transverse charge current via the z-spin to charge conversion in the PASO device. As such, the charge-spin-charge conversion process above can be modeled as a current controlled current source ($I_{out} = \eta m_z I_{supply}$). Here, $m_z = \pm 1$ is the perpendicular magnetic state

and η is the conversion efficiency. In this case, the conversion efficiency is determined by:

$$\eta = |I_{out}/I_{supply}| = \left|\frac{\Delta V}{2R_T}/I_{supply}\right| = \frac{\Delta R}{2R_T}$$

where $\Delta R$ is defined as the difference in Hall resistances between two different magnetization configurations of FGT and $R_T$ is the transverse resistance of PASO device. Here, the magnitude of $\Delta R$ is $79 \pm 9$ mΩ shown in Fig. 3B-C and $R_T$ is about 45 Ω with $\rho_{WTe_2} = 313.1$ μΩ·cm and $\rho_{FGT} = 277.0$ μΩ·cm at 80 K. In this case, we can obtain the conversion efficiency $\eta \approx 7.8 \times 10^{-4}$ with the width (w = 1.5 μm) of crossover region. To quantitatively exclude the influence of device size, we use the one-dimensional (1D) spin diffusion model reported in the previous report to characterize the conversion efficiency (*53*):

$$\eta = P_{FGT}\theta_{SH}\lambda_{WTe_2} \frac{1 - \frac{1}{\cosh\left(\frac{t_{WTe_2}}{\lambda_{WTe_2}}\right)}}{\tanh\left(\frac{t_{WTe_2}}{\lambda_{WTe_2}}\right) + \frac{\lambda_{WTe_2}\rho_{WTe_2}}{\lambda_{FGT}\rho^*_{FGT}} \tanh(t_{FGT}/\lambda_{FGT})} \frac{1}{w},$$

where $\rho^*_{FGT} = \rho_{FGT}/(1 - P^2_{FGT})$ is effective resistivity of FGT, $P_{FGT}$ is spin polarization of FGT, $\theta_{SH} = J_s/J_c$ is charge-spin conversion efficiency, $\lambda_{WTe_2,FGT}$, $\rho_{WTe_2,FGT}$ and $t_{WTe_2,FGT}$ are the spin diffusion length, the resistance and thickness of WTe$_2$ and FGT, respectively. Based on the above equation, we can obtain a parameter $\lambda_\eta = \eta w$ which is the product of the conversion efficiency and the width of crossover region, and independent of the device size. The value of $\lambda_\eta$ in our PASO device is 1.17 nm, which is one order of magnitude larger than that (0.065) reported in CoFe/Pt devices (*12*).

## The logic operations in the half adder

The logic function of half adder based on the PASO logic devices is demonstrated through circuit simulations, in which the symmetric write (magnetization switching) and read-out (charge-spin-charge conversion) operations have been incorporated. As analysis above, the read-out operation with the charge-spin-charge conversion process can be modeled as a current controlled current source ($I_{out} = \eta m_z I_{supply}$). Here, we set the conversion efficiency $\eta$ to 0.5. The process of magnetization switching in the write operation can be solved by micromagnetic simulation (see **Micromagnetic simulation** in Methods). The resulting magnetization state $m_z$ is the coefficient of the output current ($I_{out} = \eta m_z I_{supply}$) of the current source, which drives magnetization switching in the write operation of next stage. Fig. 4F demonstrates the schematic circuit of the half adder based on four PASO devices. To operate the devices in the circuit, two input signals (represented by $I_A$ and $I_B$) are used as the input signals, which are fed into Device 1 and Device 2. Here, we set the positive current pulse (+8 mA) as input logic state '1' and negative current pulse (-8 mA) as input logic state '0'. The control signals (represented by $I_{control}$) in the devices are programed to implement particular logic functions, which are labeled in Fig. 4F. The interconversion signals (represented by $I_{supply}$) are employed to read out the preceding logic stage as the charge current (modeled as the current-controlled current source) to drive the next logic stage. Fig. S12B shows the logic operations of the half adder and the operating waveforms used for control and interconversion signals on the different devices. When feeding two signals ("01" or "10") into Device 1 and Device 2, the output logic states "0" and "1" are generated from Device 3 and Device 4, respectively. Changing the input signals ("00") into Device 1 and Device 2, the output logic states "0" and "0" are generated, respectively. As such, with the results shown in Fig. 4G and Fig. S12B, we constructed the half adder circuit based on four PASO devices, which consists of AND and XOR gates to generate CARRY and SUM signals (Fig. S12A), respectively.

## Energy calculation and scalability for PASO logic device

Fig. S14A shows the schematic of the PASO logic device, in which symmetric write and read operations are used. To model the read and write operations in the PASO device, an equivalent circuit is used, as shown in Fig. S14B. First, the charge current $I_{\text{supply}}$, which flows through the vertical resistance ($R_v$), generate an output charge current $I_{\text{out}}$ through the spin-to-charge conversion, which can be modeled as a current controlled current source ($I_{\text{out}} = \eta m_z I_{\text{supply}}$) with an internal resistance $R_{\text{SCC}}$. Here, $m_z = \pm 1$ is the perpendicular magnetic state and $\eta$ is the conversion efficiency, which will be discussed later. Next, the output charge current flowing through the interconnect resistance ($R_N$) can be directly fed into another PASO material to drive its perpendicular magnetization switching. As such, the interconversion energy ($E_{\text{int}}$) and magnetization switching energy ($E_{\text{sw}}$) account for the total switching energy ($E_{\text{PASO}}$) of operating the PASO logic device.

The magnetization switching energy can be given by $E_{sw} = \int i^2 \, r dt = I_{sw}^2 R_N \tau$, where $I_{sw}$ is the critical current for switching magnetization, and $\tau$ is the pulse duration time. Since $I_{\text{out}}$ generated by the spin-charge conversion is required to exceed the critical current $I_{sw}$, the interconversion energy can be estimated by $E_{int} = \int i^2 \, r dt = I_{supply}^2 R_v \tau$, which can be reduced to

$$E_{int} = I_{sw}^2/\eta^2 R_v \tau = \frac{R_v}{R_N} E_{sw}/\eta^2.$$

Thus, the total switching energy is given by:

$$E_{PASO} = E_{int} + E_{sw} = \left(1 + \frac{R_v}{\eta^2 R_N}\right) I_{sw}^2 R_N \tau.$$

Here, we choose a thermal stability factor ($\Delta = 45$) to satisfy the industry standard retention time of ten years, and the critical current used for switching the magnetization can be calculated as (*18, 54, 55*) $I_{sw} = \frac{4e\alpha\Delta}{\hbar T_{int}\theta_{SH}} k_B T = 11.3$ uA, which is independent of the size of device. The values of material parameters and corresponding meanings are listed in Table S1 in Supplementary Materials. The vertical ($R_v$) and interconnect ($R_N$) resistances can be estimated to $90\ \Omega$ and $10\ \Omega$ with the material parameters listed in Table S1, respectively, when the device size is shrunk to 10 nm. Moreover, the current pulse duration is set to 0.5 ns by considering the ultrafast switching based on the joint effect of conventional and unconventional charge-spin conversion (*56, 57*).

The actual energy consumption of cascadable logic device depends on the efficiency of ISHE in the upper-level device, the efficiency of SHE in the next-level device and the spin current transparency. On one hand, the conversion efficiency of ISHE ($\eta = P_F \theta_{SH} \lambda_N \frac{1 - \frac{1}{\cosh\left(\frac{t_N}{\lambda_N}\right)}}{\tanh\left(\frac{t_N}{\lambda_N}\right) + \frac{\lambda_N \rho_N}{\lambda_F \rho_F^*}\tanh(t_F/\lambda_F)} \frac{1}{w_F}$) mainly depends on the spin hall angle $\theta_{SH}$ (*53*), and secondarily relies on spin polarization of ferromagnet (FM) $P_F$ and spin diffusion length of non-magnetic material(NM) $\lambda_N$. Here, $\rho_F^* = \rho_F/(1 - P_F^2)$ is effective resistivity of FM, $\lambda_{N,F}$, $\rho_{N,F}$ and $t_{N,F}$ are the spin diffusion length, the resistance and thickness of NM and FM, respectively. On the other hand, the conversion efficiency of SHE ($\xi = T\theta_{SH}$) mainly depends on spin current transparency $T$ and spin hall angle $\theta_{SH}$. Here, we can set spin current transparency $T$ as 1 considering the atomically flat interfaces in the PASO device (*35*). Thereby, the conversion efficiencies of ISHE and SHE are both dominated by $\theta_{SH}$, which depicts the intrinsic spin to charge interconversion of NM. In our estimation, the value of ξ is one, which corresponds to an experimentally observable value. Notably, such value can be experimentally realized in the previous work (*58, 59*). Thereby, the conversion efficiency is set to 0.66 with the material parameters listed in Table S1. Based on the calculated parameters above, the estimated $E_{PASO}$ is less than 10 aJ, which can be potentially reduced by seeking the quantum materials with larger charge-spin conversion efficiency ($\theta_{SH} > 1$) (*13, 42-44*) and conductivity ($> 10^6 /(\Omega \cdot m)$) (*30, 45*).

## Acknowledgements

**Funding:** This work was supported in part by the National Natural Science Foundation of China (62122036, 62034004, 61921005, 12074176 and 61974176), the Strategic Priority Research Program of the Chinese Academy of Sciences (XDB44000000), the Fundamental Research Funds for the Central Universities (020414380203, 020414380179). Crystal growth at Rutgers was supported by the center for Quantum Materials Synthesis (cQMS), funded by the Gordon and Betty Moore Foundation's EPiQS initiative through grant GBMF6402, and by Rutgers University.

**Author contributions:** F.M., B.C. and S.-J.L. conceived the idea and supervised the whole project. L.W. and J.X. fabricated the devices and performed transport measurements. Y.D., C.P., T.C, X.L, P.W., M.C., S.Y. and Z.L. assisted the measurements. L.W., J.X., B.C. and S.-J.L. analyzed the data. F.W. and H.-J.Z. carried out theoretical calculations. J.-J.X. and Y.-G.S. grew $WTe_2$ bulk crystals. X.X. and S.-W.C. grew $Fe_3GeTe_2$ bulk crystals. Z.-L.W. helped to perform Raman measurement. L.W., J.X., B.C., S.-J.L. and F.M. wrote the manuscript with input from all authors.

**Competing interests:** The authors declare no competing interests.

**Data and materials availability:** All data needed to evaluate the conclusions in the paper are present in the paper and/or the Supplementary Materials.

# Figures

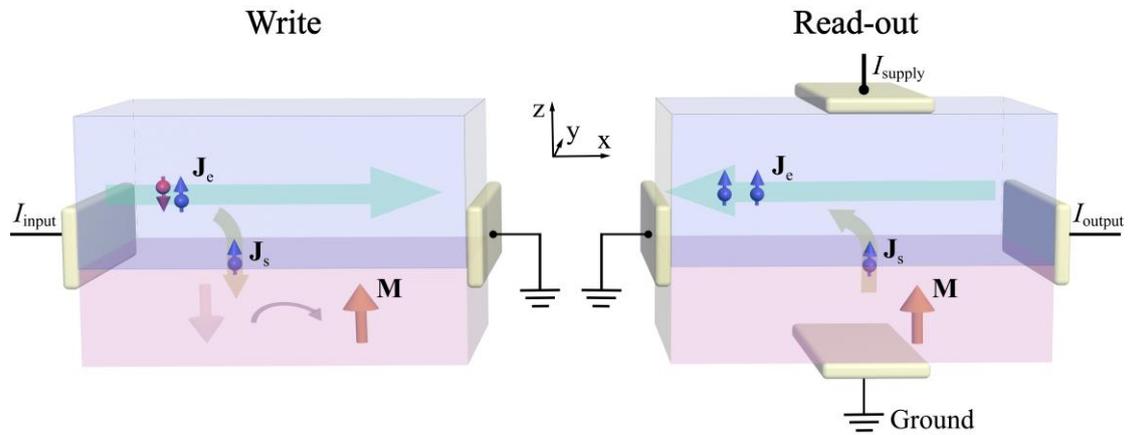

**Fig. 1. Schematic of the mechanisms for symmetric write and read-out operations.** Schematic of the write and read-out operations in the artificially designed PASO quantum material when the applied charge current is along the -x direction. The orange arrow represents the perpendicular magnetization. The green arrow represents the electron flow (along the x direction), which generates out-of-plane spin polarization accumulating at the interface and diffusing into the magnetic layer. The current-induced spin polarization pointing in the z (-z) direction is indicated by the blue (red) arrows.

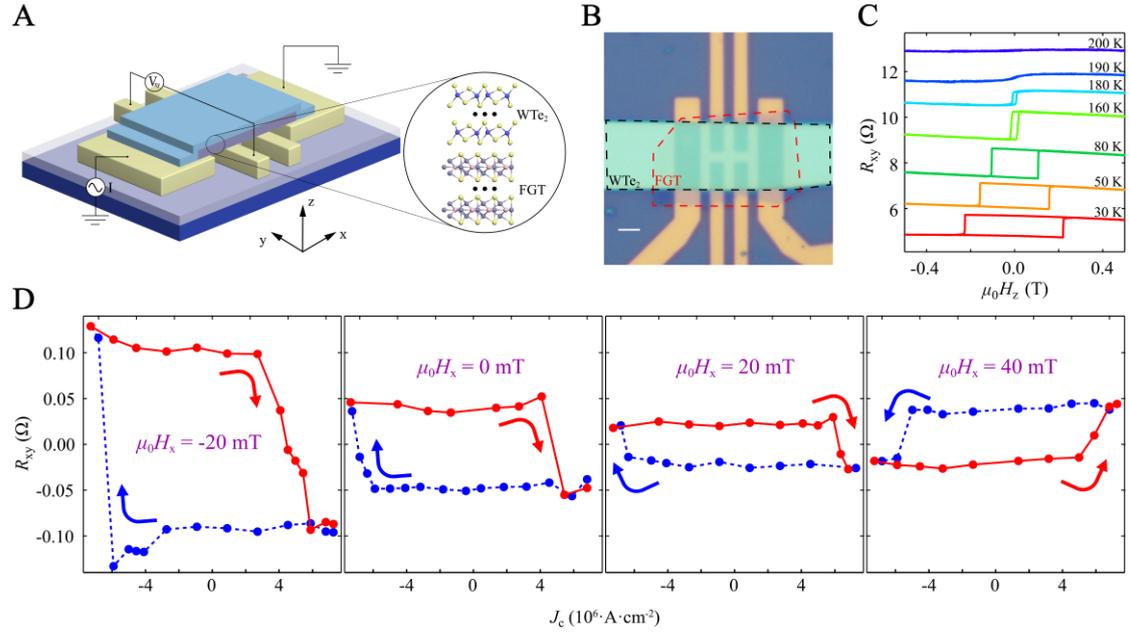

**Fig. 2. Measurement configuration and current-induced magnetization switching.** **A,** Schematic of the PASO device consisting of WTe$_2$ and FGT flakes for electric transport measurements. The applied current is along the x-axis (i.e., the a-axis of WTe$_2$). **B,** Optical image of a typical device. The scale bar is 2 μm. **C,** Hall resistance as a function of the perpendicular magnetic field for temperatures ranging from 30 to 200 K. **D,** Current-induced magnetization switching under different in-plane magnetic fields $\mu_0 H_x$ = -20 mT, 0 mT, 20 mT, and 40 mT at 120 K. The sign of Hall resistance is reversed by sweeping the pulsed current. In this PASO device, applying a current of 1 mA corresponds to a current density of $4.55 \times 10^5$ A cm$^{-2}$ in WTe$_2$.

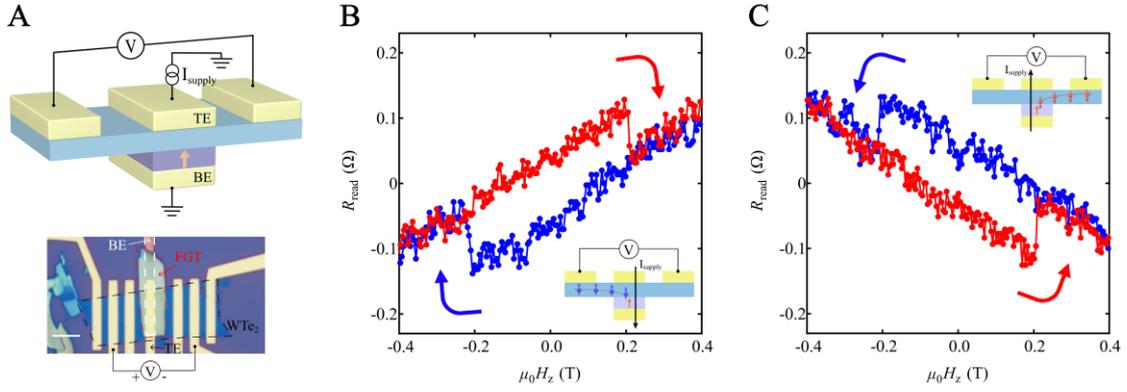

**Fig. 3. Symmetric read out of perpendicular magnetic state. A,** Schematic and optical image of the $WTe_2$/FGT device for symmetric read out of the perpendicular magnetization state. The scale bar is 5 μm. The applied current flows normal to the interface from the top electrode (TE) to the bottom electrode (BE). **B-C,** Transverse resistance $R_{read}$ as a function of perpendicular magnetic field when applying a positive **(B)** or negative **(C)** charge current. $R_{read}$ is defined as the transverse voltage ($V_{read}$) divided by the absolute current ($I_{supply}$). Inset: Schematic of the relationship between the electron flow direction and the voltage polarities under positive $H_z$. The orange arrow shows the UP state of perpendicular magnetization. The direction of the applied current is indicated by the black arrow.

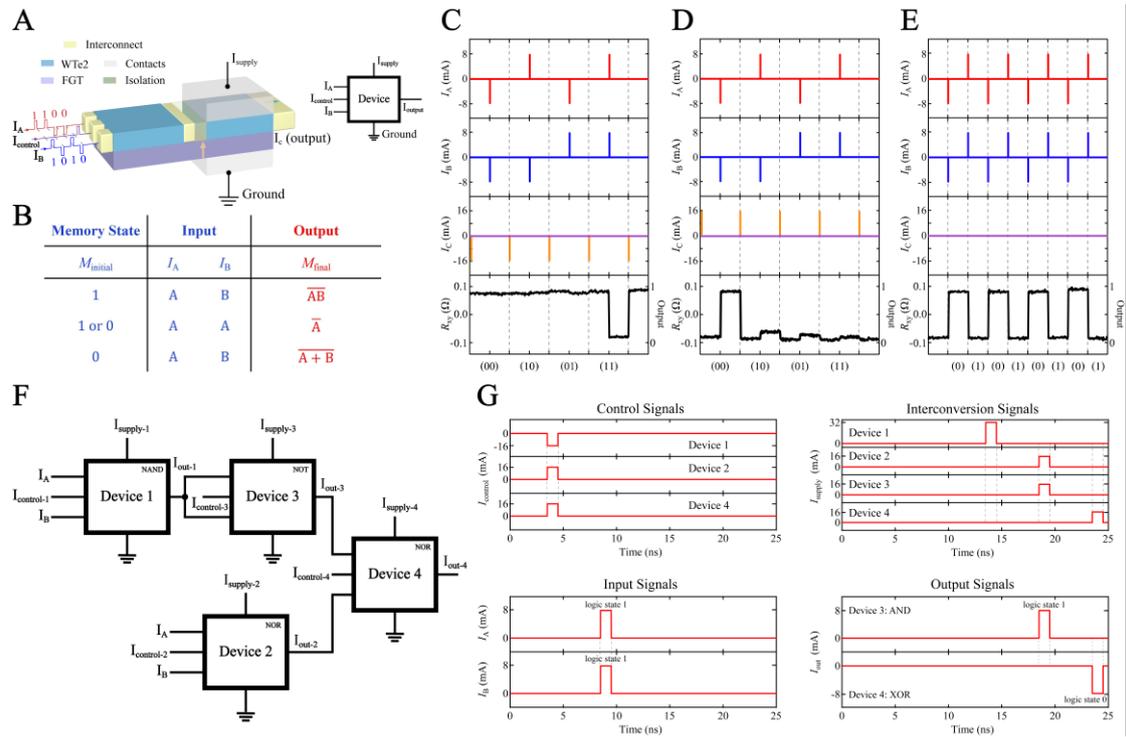

**Fig. 4. Cascaded spin-orbit logic in memory based on symmetric write and read-out operations. A,** Schematic of the PASO logic device. **B,** Logic functions of the proposed device, in which the logic output depends on the memory state represented by the magnetization state of $M_{initial}$. **C,** NAND logic function. **D,** NOR logic function. **E,** NOT logic function. **F,** Schematic of the half-adder consisting of four cascaded PASO logic devices. The logic function of each device programmed by the control signal is marked in the corresponding square. **G,** Operating waveforms of the half-adder when the logic states of two input signals are both '1'.